\documentclass[prl,twocolumn,superscriptaddress,showpacs,floatfix]{revtex4}
\usepackage{graphicx}
\usepackage{bm}% bold math
\usepackage{dcolumn}
\usepackage{amsmath}
\usepackage{amssymb}

\begin{document}

\title{Fano blockade by a Bose-Einstein condensate in an optical lattice}

\author{Rodrigo A. Vicencio}

\affiliation{Max-Planck-Institut f\"ur Physik komplexer Systeme,
D-01187 Dresden, Germany}

\author{Joachim Brand}

\affiliation{Centre of Theoretical Chemistry and Physics, Institute of
  Fundamental Sciences, Massey University, Auckland, New Zealand}

\author{Sergej Flach}

\affiliation{Max-Planck-Institut f\"ur Physik komplexer Systeme,
D-01187 Dresden, Germany}

\date{\today}

\begin{abstract}
We study the transport of atoms across a localized Bose-Einstein
condensate in a one-dimensional
optical lattice.
For atoms scattering off the condensate
we predict total reflection as
well as full transmission 
for certain parameter
values
on the basis of an exactly solvable model.
The findings of analytical and numerical calculations are
interpreted by a tunable Fano-like resonance and may lead to interesting
applications for blocking and filtering atom beams.

\end{abstract}
\pacs{03.75.Nt, 42.25.Bs, 05.60.Gg, 05.45.-a}

 \maketitle

%%%%%%FIG1
\begin{figure}[t]\centerline{\scalebox{0.9}{\includegraphics{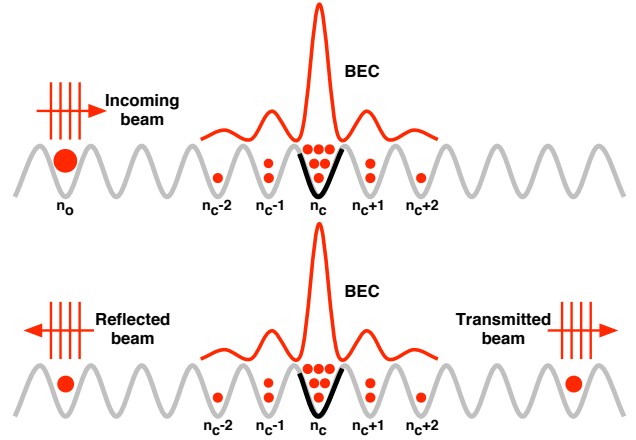}}}
\caption{(Color online). Scattering scheme in an optical lattice. The
incoming, reflected, and transmitted beams of atoms are represented as
plane waves. The BEC is centered in $n=n_c$, the only nonlinear
lattice site.}
  \label{fig0}
\end{figure}

An understanding of the transport properties of ultra-cold atoms is
vital for the development of technological applications in the fields
of matter-wave interferometry \cite{schumm05natphys} or quantum
information processing with neutral atoms
\cite{Jaksch99prl,Brennen99prl,pachos03prl,kay06pra}. Over the  
last couple of years it has been shown that optical lattices,
generated by counter-propagating laser beams and providing a periodic
potential modulation for the atoms, introduce many interesting and
potentially useful effects by modifying single atom properties and
enhancing correlations between atoms \cite{morsch06rmp}. Here we
discuss the scattering of atoms across a localized Bose-Einstein
condensate (BEC) in an optical lattice. We find that dramatic effects
of scattering resonances with either full transparency or total
reflection can occur.

Previously, transparency effects have been conjectured for the
scattering of He atoms on a film of superfluid helium-4
\cite{halley93} and similar effects have been predicted for the
scattering of atoms on a BEC in a trap of finite depth
\cite{wynveen00}. These effects were first attributed to the coherent
interactions within the target and with the scattering atoms but a
full understanding of the numerical results was not achieved. More
numerical  results were produced later \cite{poulsen03} and a Levinson
theorem was proved on general grounds \cite{brand03prl} without
revealing the mechanism for transparency effects.

In this Letter we present and analyse a very simple and analytically
solvable one-dimensional model of atom scattering by a BEC.  The
model shows transparency as well as blockade of atoms by total
reflection, which is interpreted as a Fano resonance. In the problem
studied by Fano, the resonance may both enhance and suppress
scattering due to interference \cite{Fano61pr}. 
Here the atom-atom interaction leads to an effective nonlinearity.
It was shown recently \cite{flach:084101}, that nonlinearity generates
several scattering channels which can lead to destructive interference
and Fano resonances similar to the original Fano problem. Proposed applications
in nonlinear optics and Josephson junction networks encounter various
difficulties due to inhomogeneities and dissipation 
\cite{flach:084101-b}. They are
absent in the present study, thus making the resonant atom-BEC scattering
ideal for harvesting on Fano resonances. 

\begin{figure}[t]\centerline{\scalebox{0.6}{\includegraphics{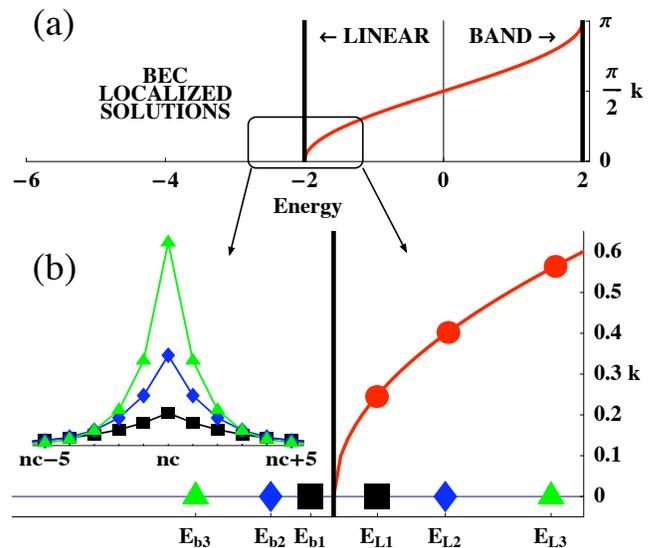}}}
\caption{(Color online). (a) Energy diagram for localized and
extended solutions. In the linear band, $E_k$ is plotted. (b) Zoom
of the region $E\sim-2$. Boxes, diamonds, and triangles correspond to
the BEC ($E_b<-2$) and to the local mode ($E_L>-2$) solutions for
$g_1$, $g_2$, and $g_3$, respectively. The corresponding BEC profiles
are shown. The filled circles correspond to the resonance condition
(\ref{res}).}
  \label{fig1}
\end{figure}

We consider a
BEC on a lattice, where interactions between atoms are present on one
lattice site only (see Fig.~\ref{fig0}). Such a situation could be
realized experimentally by combining optical lattices with atom-chip
technology \cite{haensel01prl:BEConachip,Ott01prl:BECinMicrotrap} or
in optical micro-lense arrays \cite{dumke02prl} where the $s$-wave
scattering length of atoms can be tuned by an inhomogeneous magnetic
\cite{Tiesinga93pra,Inouye1998a} or laser field
\cite{Fedichev1996a,theis04prl:OpticalFR}.
Specifically we consider the discrete nonlinear Schr\"odinger (DNLS)
equation, a classical variant of the Bose-Hubbard model appropriate
for a BEC in a periodic potential in the tight binding
limit \cite{morsch06rmp}. With interactions being present only on site
number $n_c$, we write in dimensionless form
\begin{equation}
i \frac{\partial \Psi_{n}} {\partial t} =
-\left(\Psi_{n+1}+\Psi_{n-1}\right) -\gamma |\Psi_{n_c}|^2 \Psi_{n_c}
\delta_{n,n_c},\label{eq}
\end{equation}
where $\Psi_n(t)$ is a complex amplitude of the BEC field at site $n$
and
$- \gamma=U/J$ is the interaction strength on site $n_c$. This simple
model reflects generic features of BECs in a one-dimensional optical
lattice with inhomogeneous scattering length. Furthermore, this model
could be realized quantitatively in a deep optical lattice with tight
transverse confinement \cite{smerzi03pra:nonlinearTBA}.
 For atoms with
mass $M$ in a lattice with spacing $d$ in the tight binding limit,
$J\approx 4 s^{3/4} e^{-2\sqrt{2}}/\sqrt{\pi} E_r$ is the energy scale
for tunneling between the lattice sites, where $s = V_0/E_r$ is the
depth of the optical lattice $V_0$ measured in units of the recoil
energy $E_r = 2 \hbar \pi^2/(d^2 M)$. The on-site interaction energy
per atom is $U= 4\pi a_s \hbar^2
\int d^3x |\psi({\mathbf x})|^4/M$, where $a_s$
is the tunable $s$-wave scattering length at the nonlinear site $n_c$ 
and $\psi({\mathbf x})$ is
the localized Wannier state associated with the lowest Bloch band of
the lattice. 
The number of atoms in the lattice
is given by $N=\sum_n |\Psi_n|^2$.
Small-amplitude plane-wave solutions of Eq.~(\ref{eq}) take the form 
$\Psi_n=\Psi_0 \exp (i k
n) \exp (-i E_k t)$ and satisfy the relation
\begin{equation}
E_k\equiv-2\cos k,
\label{Ek}
\end{equation}
which defines the band of single-particle energies $E_k \in [-2,2]$
[see Fig.\ref{fig1}(a)]. The unit of dimensionless energy is $J$ and
the quasi-momentum $k$ is measured in units of $d^{-1}$.

First, we look for localized and stationary solutions of
Eq.(\ref{eq}), corresponding to the BEC centered in $n=n_c$. For
simplicity we assume that interactions are attractive and thus
$\gamma>0$. As an
ansatz, we take an exponentially localized profile:
 $\Psi_n(t)= b_n
(t)=b_{n_c} x^{|n-n_c|} \exp (-i E_b t)$, where $b_{n_c}$ is the
condensate amplitude, $|x|<1$, and $E_b$ is the respective
energy. Inserting this expression into (\ref{eq}), we obtain that
\begin{equation}
E_b=-\sqrt{4+g^2}\ \ \text{and}\ \ x=-(E_b+g)/2, \label{bec}
\end{equation}
where $g\equiv\gamma b_{n_c}^2$ ($g>0$). Eqs.~(\ref{bec}) correspond
to solutions for localized BECs with $E_b$ being outside of the
band $E_k$ 
[$E_b<-2$, see Fig.\ref{fig1}(a)]. They are similar to bright
lattice solitons pinned to the nonlinear lattice site. These localized
modes exist only above a threshold $N_b=-E_b/\gamma > N_b^{th}$ 
\cite{Tsironis94pre}, given by
$N_b^{th}=2/\gamma$.
We assume that $N_b$ is significantly larger than the threshold, which
should be easily achieved in a possible experiment.

By controlling the number of atoms in the BEC or by tuning the
nonlinear coefficient, we can easily modify the BEC energy
$E_b=-\gamma N_b$ (this is equivalent to modifying the parameter
$g$). This is one of the keys for a tunable Fano-blockade scheme. As
we will show later, this energy is directly related to the energy
where zero transmission of the atom beam through the BEC is observed.

%
%%%%%%FIG2
We consider three different values for $g$ as
examples: $g_1=0.36$, $g_2=0.6$, and $g_3=0.9$. The corresponding
energies [$E_{b1}=-2.03$ (box), $E_{b2}=-2.09$ (diamond), and
$E_{b3}=-2.19$ (triangle)] and the profiles of the BEC states 
are shown in the left
part of Fig.~\ref{fig1}(b). By doing a linear stability analysis,
we find that
all these solutions centered in $n=n_c$ are stable. This is important
for two reasons. First, experimental generation of these
states should be possible.
Second, the scattering of small-amplitude beams by these modes
can be viewed as a small perturbation of the latter, which will stay
small if the BEC is stable, but would grow (and the BEC would decay)
otherwise.

In order to study the scattering of a 
propagating atom beam by the
localized BEC, we consider:
\begin{equation}
\Psi_n(t) = \phi_n(t)+b_n (t),
\label{P+N}
\end{equation}
where $\phi_n\ll b_n$. We linearize Eq.(\ref{eq}) with respect to
$\phi_n(t)$, obtaining the following equation for the atom beam:
\begin{equation}
i \frac{\partial \phi_{n}} {\partial t} =-(\phi_{n+1}+\phi_{n-1}) -g
(2\phi_{n_c}+e^{-2iE_b t}\phi_{n_c}^*) \delta_{n,n_c}\ .
\label{pw}
\end{equation}
Far away from $n=n_c$, the solution of this equation corresponds to a
propagating plane wave which satisfies (\ref{Ek}). The
localized BEC generates a scattering potential for the atom beam which
has a constant and a time-dependent part. Equation (\ref{pw}) is
similar to the one found in Ref.\cite{flach:084101} for the scattering of
plane waves against a discrete breather. In that work 
the appearance of Fano resonances \cite{Fano61pr} was studied in a
full nonlinear lattice in the approximation of a very localized
nonlinear mode. The scattering potential is then reduced to one
lattice site and implies $g \gg 1$, which makes the discrete breather
a very strong scatterer for almost all incoming plane waves.
Our setup allows us to tune $g$ to smaller
values, and thus admits Fano resonances on a background of almost
perfect transmission. 

We solve the equation (\ref{pw}) by using a Bogoliubov
transformation \cite{poulsen03} given by:
\begin{equation}
\phi_n(t) = u_n e^{-i E t}+v_n^* e^{-i(2E_b-E)t}\ ,
\label{bogo}
\end{equation}
and by inserting it in (\ref{pw}) we obtain the discrete Bogoliubov
(DB) equations:
\begin{eqnarray}
E u_n=-(u_{n+1}+u_{n-1})-g (2u_{n_c}+v_{n_c})\delta_{n,n_c},\label{coeq1}\\
(2E_b-E) v_n=-(v_{n+1}+v_{n-1})-g (2v_{n_c}+u_{n_c})\delta_{n,n_c}.
\label{coeq2}
\end{eqnarray}
Here $u_n$ corresponds to an {\it open} channel which, far away from
$n_c$, represents a propagating atom beam for which the energy is in
the band $E=E_k \in [-2,2]$. Contrary, $v_n$ represents a {\it closed}
channel whose extended states far away from the scattering center have
$2E_b-E \notin [-2,2]$. They are thus located outside the open channel
continuum and cannot be excited in the same energy range.  However, even in
the absence of any coupling between both channels, the scattering
center provides also a localized state in the spectrum of the closed
channel with energy $E_L$. The localized state, by definition, is
located outside the band of extended states in the $v_n$ channel, but
may be located inside the $u_n$-channel band $E \in [-2,2]$. In such a
case, taking the coupling between channels into account, we encounter
a Fano resonance for $E_L=E_k$.

Let us consider the situation when
both channels are decoupled. For this particular situation, the closed
channel equation is given by
\[
(2E_b-E) v_n=-(v_{n+1}+v_{n-1})-2g v_{n_c}\delta_{n,n_c},
\]
and admits a localized solution: $v_n=v_{n_c}\ w^{|n-n_c|}$ ($|w|<1$), for which
\[
w=-g+\sqrt{1+g^2}\ \ \text{and}\ \ E=E_L\equiv2(E_b+\sqrt{1+g^2}).
\]
We call this solution the local mode (LM). $E_L$ corresponds to the
LM energy, which is always inside the continuum of the open channel: 
if $g\rightarrow 0
\Rightarrow E_L\rightarrow-2$ and, if $g\rightarrow \infty\Rightarrow
E_L\rightarrow0$. Therefore, due to the time dependence of the original
scattering potential, the closed channel is able to resonate with the
open one at a frequency that depends on externally
controllable parameters.

Keeping in mind the localized nature of the LM and the propagating one
of the open channel, we make the
following ansatz:
\begin{eqnarray}
u_n=\left\{\begin{array}{c} a_1\ e^{ik(n-n_c)}+b_1\ e^{-ik(n-n_c)}\ ;\
n<n_c \\ c_1\ e^{ik(n-n_c)} \hspace{2.6cm} ;\ n\geq n_c \\
\end{array}\right.,\label{sca1}\\ v_n=\bar{v}_{n_c}\
\bar{w}^{|n-n_c|}.\label{sca2}
\end{eqnarray}
Here, $a_1$, $b_1$ and $c_1$, represent the incoming, reflected and
transmitted beam amplitudes, respectively. $\bar{v}_{n_c}$ correspond
to the closed channel amplitude and $|\bar{w}|<1$. The beam
quasimomentum $k$ can be generated in the experiment by using a
phase imprinting method \cite{Denschlag2000a}, Bragg scattering, or
simply by acceleration of the matter-wave probe in an external potential.
We solve analytically the scattering problem by inserting (\ref{sca1})
and (\ref{sca2}) in (\ref{coeq1}) and (\ref{coeq2}) for
$n=n_c,n_c\pm1$. By doing so, we obtain that the open channel
satisfies (\ref{Ek}), $a_1+b_1=c_1$, $\bar{w}=w$, and that the
transmission is given by:
\begin{equation}
T(k)=\frac{4\sin^2 k}{4\sin^2 k+
\left(2g+\frac{g^2}{\sqrt{(E_k-2E_b)^2-4}-2g}\right)^2}
\label{T}
\end{equation}
($T\equiv |c_1/a_1|^2$). Resonances occur when the denominator
diverges or when $\sqrt{(E_k-2E_b)^2-4}-2g=0$. 
The condition for the resonance is
\begin{equation}
E_k= E_L \Rightarrow k_L \equiv \arccos (-E_L/2).
\label{res}
\end{equation}
Eq.(\ref{res}) implies that the
transmission for atom beams through the BEC is reduced to zero when a
LM is generated in the process, i.e. when the closed channel resonates
with the open one.
%
%%%%%%FIG3
\begin{figure}[tbp]\centerline{\scalebox{1}{\includegraphics{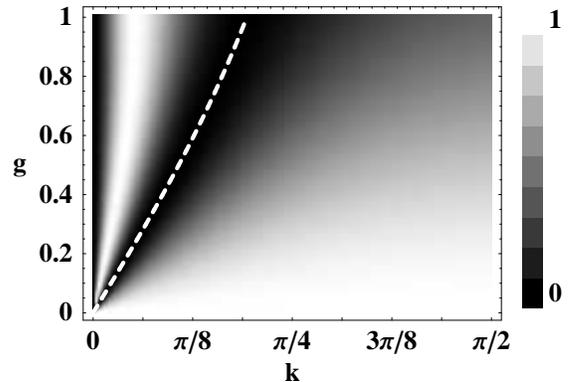}}}
\caption{$T$ versus $k$ and $g$. The dotted white line correspond to
$T(k_L)=0$, the resonance position.}
  \label{fig2}
\end{figure}

In Fig.\ref{fig1}(b), we show the corresponding energies for the
LM's, for the three values of $g$, $E_{L1}=-1.94$ (box),
$E_{L2}=-1.84$ (diamond), and $E_{L3}=-1.69$ (triangle). The curve in
that figure corresponds to Eq.~(\ref{Ek}), and the filled circles
correspond to the value of $k$ for which the resonance is expected
(\ref{res}). In principle, any energy in the interval $\{-2,0\}$ can
be a good candidate for the observation of a Fano resonance in this
setup. But, as we will show later, the response of the system is not
always the same and it essentially depends on the BEC profile.

\begin{figure}[tbp]\centerline{\scalebox{0.93}{\includegraphics{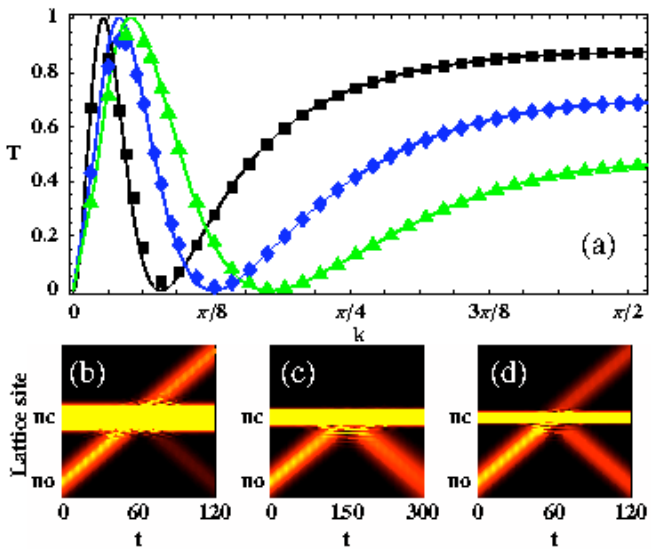}}}
\caption{(Color online). (a) $T$ versus $k$ for the three values of
$g$. Lines: eq.~(\ref{T}), points: real time numerical simulations of
eq.~(\ref{eq}) for $g_1$ (boxes), $g_2$ (diamonds), and $g_3$
(triangles). (b)--(d) Evolution of $|\Psi_n(t)|^2$ in space 
and time: (b) $g_1, k=1.37$, (c)
$g_2, k=0.39$, (d) $g_3, k=1.37$.}
  \label{fig3}
\end{figure}

We compute the transmission 
$T=T(k,g)$ and we vary the beam velocity $k$ and the nonlinear
parameter $g$. Fig.\ref{fig2} shows this sweep of parameters, where
dark and bright regions represent a low ($T\rightarrow0$) and a high
($T\rightarrow1$) transmission response, respectively. In
Fig.\ref{fig3}(a) we show some $T(k)$ for three values of
$g$. As $g$ increases the width and the position 
(dashed white line in Fig.~\ref{fig2}) of the resonance increase. Thus,
the more localized the BEC becomes, the stronger
it reflects the atom beam off resonance. As expected from
(\ref{res}), increasing $g$ (which corresponds to decreasing
$E_b$ or increasing $N_b$) leads to an increase of the resonance energy. By tuning
the nonlinear parameter $g$, we can thus choose the
amount of the beam which passes
through the BEC. Off resonance (for larger values of $k$),
we can select the percentage of the incoming beam that is transmitted
for a defined quasimomentum. Therefore, the actual setup can be used
as a $100\%$ blockade or as a selective filter.
%
%%%%%%FIG4

Now, we look for a numerical confirmation of our theoretical
description for the scattering of an atom beam against a localized
BEC. In the simulations, we initialize the atom beam with a Gaussian profile:
$\phi=\phi_0 \exp[-\alpha (n-n_0)^2]\exp [ik(n-n_0)]$ with
$\phi_0=0.01$ and $\alpha=0.001$. $n_0$ is the initial location of the
center of the distribution and well separated from the BEC to avoid an initial
interaction. $k$ is the initial quasimomentum. The amplitude
$\phi_0$ was chosen to be very small compared to the BEC amplitude
($\phi_0/b_{n_c}\sim 1\%$) in order to justify
(\ref{pw}). The value of $\alpha$ implies a spatial width of approximately
$60$ sites and a reciprocal width in $k$-space of $0.12$. 
With this choice we can
clearly observe the resonant response of the system. In
Fig.\ref{fig3}(a) the symbols denote our numerical results for
three different values of $g$. The agreement
between theory and simulations is almost perfect. We have some
disagreement for small values of $k$ where the group velocity is too
small and the computation of $T$ is more complicated. Figs.\ref{fig3}
(b), (c), and (d) show some numerical examples of the scattering
process with a transmission of $80\%$, $0\%$ and $40\%$, respectively. It is worth mentioning that, if we extend the nonlinearity to three sites of the array, similar transmission curves are observed. This shows the robustness of our theoretical results in a more realistic experimental setup.

In conclusion, we have investigated Fano resonances in the context of Bose-Einstein condensates in an optical lattice. The implementation of this idea can be viewed as a powerful tool for controlling the transmission of matter waves in interferometry and quantum information processes. Fano resonances rely on destructive interference and are thus inherent to wave dynamics. An observation of these resonances in atom-BEC scattering would provide, in addition to tunable filters, a new demonstration of the quantum matter wave character of ultracold atoms.

\end{document}